\documentclass{WileyMSP-template}

\usepackage[utf8]{inputenc}
\usepackage{hyperref}
\usepackage{url}

\begin{document}
\pagestyle{fancy}
\rhead{\includegraphics[width=2.5cm]{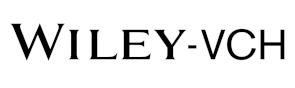}}

\title{\textbf{Enhanced Sensitivity to Blackbody Radiation in Spintronic Poisson Bolometers}}

\maketitle


\author{Ziyi Yang, Sakshi Gupta, Jehan Shalabi, Daien He, Leif Bauer, Mohamed A. Mousa, Angshuman Deka, Zubin Jacob*}




\begin{affiliations}
{\normalsize
Ziyi Yang, Jehan Shalabi, Dr. Daien He, Dr. Leif Bauer, Dr. Mohamed A. Mousa, Prof. Zubin Jacob\\
Elmore Family School of Electrical and Computer Engineering, Purdue University, West Lafayette, Indiana 47907, USA\\
\vspace{0.3em}
Sakshi Gupta, Prof. Zubin Jacob\\
Department of Physics and Astronomy, Purdue University, West Lafayette, Indiana 47907, USA\\
\vspace{0.3em}
Dr. Angshuman Deka\\
Birck Nanotechnology Center, Purdue University, West Lafayette, Indiana 47907, USA\\
\vspace{0.3em}
Email Address: zjacob@purdue.edu\\}
\end{affiliations}



\justifying
\begin{abstract}
{\normalsize

 \textbf{Abstract:} High-sensitivity long-wave infrared (LWIR) detection is crucial for observing weak thermal radiation. Recently, the Poisson bolometer has been proposed as a fundamentally new platform for uncooled infrared detection. In contrast to traditional analog detectors, where signal and noise are determined by continuous currents or voltages, the Poisson bolometer's signal and noise are governed by Poissonian counting statistics regardless of the light source. In this work, we demonstrate advancements in uncooled infrared detection towards cryogenic-level sensitivity through the integration of spintronic and plasmonic materials. Specifically, a spintronic Poisson bolometer is experimentally integrated with a plasmonic nanoantenna array optimized for broadband LWIR absorption to enhance the temperature increase of the sensing layer. The plasmonic absorber exhibits an absorptance exceeding 60\% across the LWIR spectrum, matching the peak of room-temperature blackbody radiation. We demonstrate that these devices are capable of achieving a noise equivalent differential temperature (NEDT) of 35 mK at a 50 Hz frame rate, demonstrating room-temperature performance comparable to the most sensitive uncooled LWIR detectors reported to date. This work opens up a pathway to removing bulky and expensive cooling requirements from high-sensitivity LWIR detection and imaging applications, such as remote sensing, high-speed imaging, and industrial monitoring.
}
\end{abstract}

\section{Introduction}

High-sensitivity long-wave infrared (LWIR) detectors are essential for detecting weak thermal radiation \cite{rogalski2019infrared}. In many scenarios, the detectable infrared power is low due to limited radiative power or other measurement conditions. For example, low-temperature objects (\(<\) 200 K) and low-emissivity surfaces emit faint infrared radiation \cite{salihoglu2018graphene, xu2020high, HowDoesE48:online}. Moreover, targets with small temperature differences relative to their background require high thermal contrast resolution from detectors \cite{jordan2024comparison, grimming2021ground}. Limited photon collection occurs in long-range imaging and high-speed imaging, which have high signal loss and short integration times, respectively \cite{flannigan2022mid, HighSpee43:online}. These challenges underscore the need for highly sensitive LWIR detectors, which enable applications in remote sensing of subtle geological variations \cite{rashman2020terrestrial}, detection of cold celestial bodies \cite{lawson2000brightness, wishnow1998long}, and imaging of low-emissivity \cite{HowDoesE48:online, salihoglu2018graphene} or rapidly evolving scenes \cite{landmann2019simultaneous, landmann2023high}.

State-of-the-art LWIR detectors can be classified into two types: cooled and uncooled \cite{rogalski2019infrared}. Common cooled LWIR detectors include type-II superlattice (T2SL) photodetectors typically based on InAs/GaSb or related antimonides \cite{hu2025progress, zhou2023dual, FLIRA6780SLS, lee2021comparative, jiang2025high}, quantum well infrared photodetectors (QWIP) employing GaAs/AlGaAs multiple quantum wells \cite{lu2025lwir, ivanov2023qwip}, mercury cadmium telluride (HgCdTe or MCT) photodiodes \cite{zhang2023dark, varavin2020photodiodes}, and superconducting nanowire single-photon detectors (SNSPDs) commonly fabricated from niobium nitride (NbN), niobium titanium nitride (NbTiN), or tungsten silicide (WSi) nanowires \cite{chen2025sub, lita2022development, snspd_wsi}. Though demonstrating high response speed, high spectral selectivity, and high sensitivity, commercial cooled detectors are often based on narrow-bandgap semiconductors, which are limited by thermally activated dark current and defect-mediated generation–recombination noise \cite{rogalski2003quantum, piotrowski2004uncooled, xue2023high}. These noise mechanisms grow rapidly at room temperature and therefore necessitate cryogenic cooling to achieve high sensitivity \cite{rogalski2023infrared}. Emerging cooled infrared sensors are based on superconductors, which require cryogenic temperatures due to their low critical temperatures \cite{natarajan2012superconducting}.

On the other hand, the most common uncooled LWIR detectors are microbolometers, which measure radiation via a temperature-induced change of the material’s electrical properties \cite{rogalski2019infrared}. Microbolometers typically employ vanadium oxide (VOx), amorphous silicon (a-Si), and are sometimes enhanced with metal-insulator-metal (MIM) absorbers to improve optical coupling \cite{wu2023one, FLIRA655, fonseca2025new_amorSilicon, chen2020polarization}. Thermomechanical detectors are a promising new class of uncooled detectors, which are based on a shape memory polymer (SMP) or a perforated MIM absorber membrane \cite{adiyan2019shape, das2023thermodynamically}. However, both microbolometers and thermomechanical detectors face inherent constraints such as slow thermal time constants \cite{yu2020low, rogalski2021trends}, relatively large pixel footprints \cite{yu2020low}, and susceptibility to thermal or mechanical noise \cite{vashaee2024thermoelectric, adiyan2019shape}. These limitations restrict response speed, resolution scaling, and ultimate sensitivity compared to cryogenically cooled photon detectors. As a result, achieving high sensitivity at room temperature remains an outstanding challenge for LWIR thermal detection. The most common figure of merit for infrared detector sensitivity is the noise equivalent differential temperature (NEDT), or interchangeably noise equivalent temperature difference (NETD), representing the minimum temperature difference that can be sensed by the detector with a signal-to-noise ratio (SNR) equal to one \cite{bianconi2020recent, rogalski2019infrared}. We compare the NEDT of state-of-the-art cooled and uncooled LWIR detectors in \textbf{Figure 1} (a) and (b).

\begin{figure}[t!]
\centering
  \includegraphics[width=0.8\linewidth]{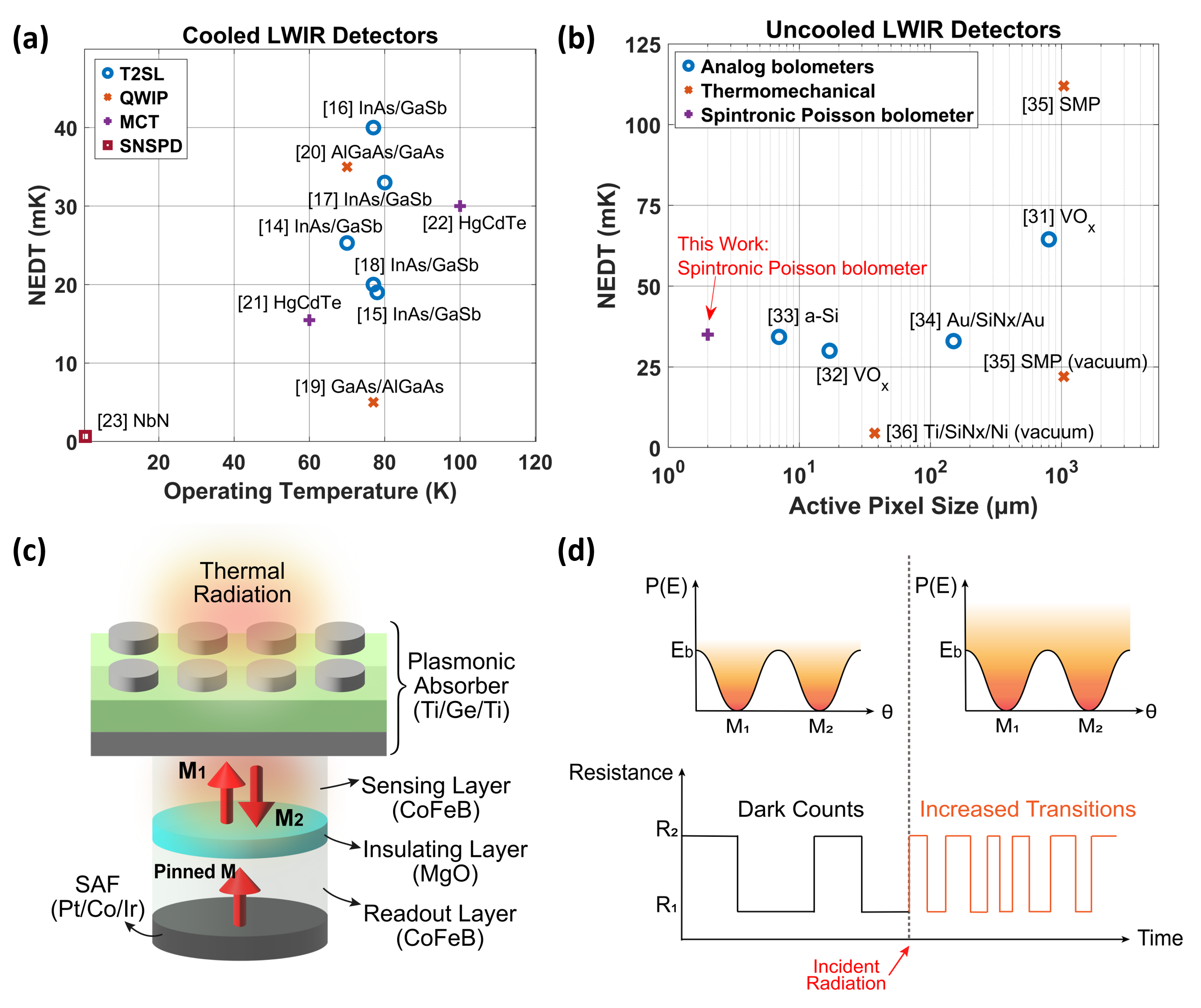}
  \caption{Achieving cooled thermal sensitivity with uncooled spintronic Poisson bolometers. (a) A comparison of state-of-the-art cooled LWIR detectors based on NEDT measured in vacuum. (b) A comparison of state-of-the-art uncooled LWIR detectors based on NEDT measured in air unless labeled “vacuum”. In this work, the best NEDT achieved with our spintronic Poisson bolometer is 35 mK at 50 Hz (measured in air), which ranks among the most sensitive uncooled LWIR detectors. (c) Spintronic Poisson bolometer device structure schematic. The device is composed of a magnetic tunnel junction (MTJ) and a plasmonic nanoantenna array on top, which acts as the absorber. The MTJ is composed of a sensing layer, an insulating layer, a readout layer, and a synthetic antiferromagnetic layer (SAF). The SAF pins the magnetization of the readout layer to the perpendicular direction, while the sensing layer has two stable states of magnetization (\(M_1\) and \(M_2\)) that are separated by an energy barrier. The device exhibits tunnel magnetoresistance (TMR) depending on the relative orientation of the magnetization of the sensing layer and the readout layer. The plasmonic absorber is optimized for broadband LWIR absorption, enabling a higher temperature increase in the sensing layer. (d) Operation mechanism of spintronic Poisson bolometers. When no light is incident, natural heat in the sensing layer causes a low rate of transitions to occur between $M_1$ and $M_2$. These transitions are read out through the device’s resistance change, which we call dark counts. When light is incident, the temperature of the sensing layer rises, increasing the rate of transitions, which we call bright counts.
}
  \label{fig1}
\end{figure}

Recently, the first room-temperature Poisson bolometer, the spintronic ultrafast nanoscale bolometer, was proposed \cite{bauer2025exploiting, mousa2025neural}. The Poisson bolometer operates in a probabilistic regime dominated by Poissonian noise, establishing a novel detection paradigm \cite{yang2025optical}. In contrast to traditional analog detectors, where signal and noise are determined by continuous currents or voltages, the Poisson bolometer has both signal and noise governed by Poissonian counting statistics regardless of the light source, with the mean count rate modulated by incident radiation. Relative to conventional analog detectors limited by continuous readout noise, a digital Poisson-counting detector can offer an advantage in low-signal and read-noise-limited regimes when incident radiation produces a measurable change in the mean event rate over the observation window. It has been demonstrated in \cite{fossum2016quanta, buchner2021analytical} that the thresholding of discrete events can effectively eliminate downstream continuous electronic noise.


The Poisson bolometer in \cite{bauer2025exploiting} was implemented using engineered spintronic materials, specifically a stochastic magnetic tunnel junction (MTJ) based on CoFeB/MgO/CoFeB and a synthetic antiferromagnetic layer (SAF). Thermally activated transitions between two discrete magnetization states were utilized to achieve a digital response to incident radiation. Their spintronic Poisson bolometer demonstrated an NEDT of 103 mK at a 25 Hz frame rate, with an array of plasmonic nanoantennas incorporated onto the transduction layer to enhance sensitivity. However, the plasmonic absorber was designed to be highly absorptive to mid-wave infrared (MWIR) light, with only limited absorption in the LWIR, thus deviating from the radiation peak of a blackbody at room temperature.

In this work, we advance uncooled infrared detection toward cryogenic-level sensitivity by integrating a spintronic Poisson bolometer (SPB) with a broadband LWIR plasmonic nanoantenna array to enhance thermal absorption. By optimizing integrated spintronic and plasmonic materials, we experimentally achieved an NEDT of 35 mK at a 50 Hz frame rate, and multiple NEDT values close to or below 100 mK, demonstrating room-temperature performance that ranks among the most sensitive uncooled LWIR detectors reported to date. The active pixel size of our spintronic Poisson bolometer is approximately 2 \textmu m by 2 \textmu m, potentially allowing for high pixel density and resolution scaling. These results represent a significant improvement over previous spintronic Poisson bolometers and push the frontier of uncooled LWIR detection towards cryogenic-level sensitivity. We believe that this work opens the way to a broad range of applications in high-sensitivity LWIR sensing and imaging, including remote sensing, high-speed imaging, cryogenic system diagnostics, and industrial monitoring.


\section{Results}

The spintronic Poisson bolometer we study here is composed of a broadband LWIR plasmonic absorber and a stochastic MTJ. This integration enables a fundamentally new approach to detecting thermal radiation with tunable, magnetically controlled electrical readout. A simplified schematic of the device structure is shown in \textbf{Figure 1}  (c). The magnetic tunnel junction is comprised of four primary layers: a magnetic sensing layer (i.e., free layer) based on CoFeB, a nonmagnetic insulating layer (i.e., tunnel barrier) based on MgO, a magnetic readout layer (i.e., fixed layer) also based on CoFeB, and a synthetic antiferromagnetic layer (SAF) consisting of Co/Pt multilayers separated by an Ir spacer. Detailed material stack information can be found in the Supporting Information. The MTJ's tunnel magnetoresistance depends on the orientation of the sensing layer’s magnetization relative to that of the readout layer. The readout layer’s magnetization is fixed by the SAF layer located directly below, which is out-of-plane. The sensing layer has two stable magnetization states (\(M_1\) and \(M_2\)) that are separated by an energy barrier. As a result, the device's overall resistance depends on the orientation of the sensing layer's magnetization. The resistance is low when the magnetization of the sensing layer and the magnetization of the readout layer are parallel; the resistance is high when they are antiparallel. Experimentally, the resistance is read out by applying a bias voltage $V_{bias}$ to the spintronic Poisson bolometer and recording the bolometer’s voltage signal using an oscilloscope.

To estimate the incident power from the measured device response, we establish a mathematical relationship between the incident power on the device and its mean count rate. When no thermal radiation is incident, the device's natural heat causes discrete transitions between the two magnetization directions, which we refer to as dark counts. When thermal radiation is incident, the plasmonic absorption layer couples the heat to the sensing layer, which increases the temperature of the sensing layer. This increases the probability of the transitions in the MTJ's sensing layer, leading to an increased count rate, as shown by \textbf{Figure 1} (d). The mean of the interarrival times of these stochastic transitions \(\tau_{M_1,(M_2)}\) are governed by the Néel-Arrhenius law \cite{coffey2012thermal}:

\begin{equation}
\label{eq:N-A law}
\tau_{M_1,(M_2)} = \tau_0 \exp\left( \frac{E_b}{k_B T} \right)
\end{equation}

\noindent where \(\tau_0\) is the attempt time, \(k_B\) is Boltzmann’s constant, $T$ is the temperature of the sensing layer, and $E_b$ is the energy barrier given by

\begin{equation}
E_b = \frac{M_S V H_k}{2}
\end{equation}

\noindent where $M_S$ is the saturation magnetization, $V$ is the volume of the sensing layer, and $H_k$ is the magnetic anisotropy field. The Néel–Arrhenius law can alternatively be written in terms of the rate of transitions, i.e., count rate:

\begin{equation}
\lambda(T) = \lambda_0 \exp\Bigg(-\frac{E_b}{k_B T}\Bigg)
\end{equation}

\noindent where $\lambda_0 = 1/\tau_0$ is the attempt rate. On the other hand, the relaxation time is known to follow an exponential distribution \cite{Hayakawa2021RTN, Adeyeye2025Exponential}, and the counts follow a Poisson distribution \cite{Garanin1998, Brown1963}. Therefore, during an observation window, or integration time $\Delta t$, the number of counts $N(\Delta t)$ is a random variable with Poission distribution:

\begin{equation}
N(\Delta t) \sim \mathrm{Poisson}(\mu)
\end{equation}

\noindent The mean of the Poisson distribution is $\mu = \lambda \Delta t$. When no light is incident, we assume the temperature of the free layer is $T_0$, then the number of counts in a time window $\Delta t$ is given by

\begin{equation}
N_0(\Delta t) \sim \mathrm{Poisson}\Big(\mu_0 = \lambda_0 \Delta t \, e^{-E_b/(k_B T_0)}\Big)
\end{equation}

\noindent According to \cite{bauer2025exploiting}, the count rate increases linearly with incident power; therefore, the temperature increase $\Delta T = T-T_0$ can be assumed to be linear with the incident power $P_{\rm in}$ when $\Delta T \ll T_0$:

\begin{equation}
T = T_0 + \eta P_{\rm in}
\end{equation}

\noindent where $\eta$ is the thermal impedance (unit: Kelvin/Watt) \cite{rogalski2019infrared}. The number of counts becomes:

\begin{equation}
N_1(\Delta t) \sim \mathrm{Poisson}\Big(\mu_1 = \lambda_0 \Delta t \, e^{-E_b/(k_B (T_0 + \eta P_{\rm in}))}\Big)
\label{num_brightcount}
\end{equation}

\noindent The mean count rate shifts from $\mu_0$ to $\mu_1$ due to the incident light, and this is our key method for estimating $P_{\rm in}$, which can be solved as:

\begin{equation}
P_{in} = \frac{1}{\eta} \left( \left[ \frac{1}{T_0} - \frac{k_B}{E_b} \ln\left(\frac{\mu_1}{\mu_0}\right) \right]^{-1} - T_0 \right)
\end{equation}

 \noindent Since $\hat{\mu_0} = N_0, \hat{\mu_1} = N_1$, by the invariance property of the maximum likelihood estimator \cite{ziegel2002statistical}, the estimator of $P_{\rm in}$ can be obtained by: 

\begin{equation}
\hat{P}_{in} = \frac{1}{\eta} \left( \left[ \frac{1}{T_0} - \frac{k_B}{E_b} \ln\left(\frac{N_1}{N_0}\right) \right]^{-1} - T_0 \right)
\end{equation}

\noindent Thus, we can estimate the incident power through changes in the counts recorded in a fixed integration time.



\begin{figure}[ht!]
\centering
  \includegraphics[width=\linewidth]{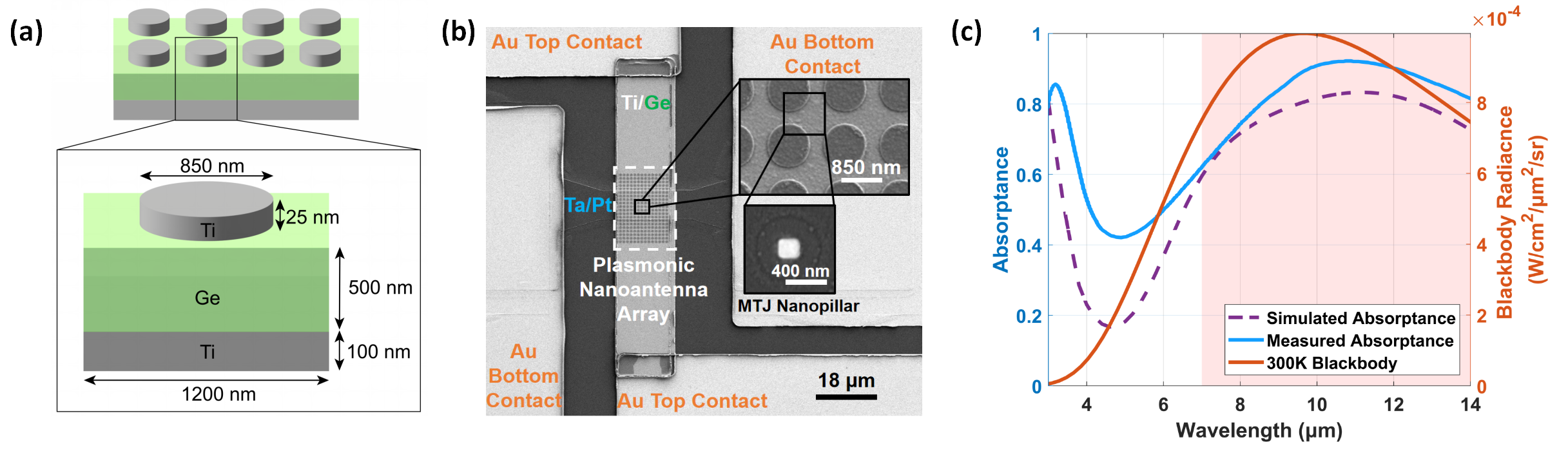}
  \caption{Design and characterization of the LWIR plasmonic nanoantenna array absorber. (a) Plasmonic absorber unit cell structure and dimensions. The absorber is a metal-dielectric-metal (MDM) metasurface with circular Ti nanoantennas on Ge-Ti layers. The dimensions are engineered to achieve enhanced broadband LWIR absorption. (b) SEM image of the fabricated spintronic Poisson bolometer. The top contact strip area above the MTJ is covered with the plasmonic absorber. The inset shows the plasmonic nanoantenna array and the MTJ nanopillar before the nanoantenna array is deposited. (c) Left vertical axis: Simulated and measured absorptance of the plasmonic absorber. The design enables broadband LWIR absorption higher than 60\% with a peak around 11 µm wavelength. Right vertical axis: Simulated blackbody spectral radiance at 300 K. At room temperature, the blackbody radiation peaks at the 7-14 \textmu m LWIR spectrum (red shaded region). The absorption of the plasmonic nanoantenna array peaks at the same region to maximize the temperature increase in the sensing layer of the MTJ.
}
  \label{fig2}
\end{figure}

 To optimize thermal absorption and enhance the temperature rise in the device’s sensing layer, we integrated a broadband LWIR plasmonic absorber onto a spintronic Poisson bolometer. The plasmonic absorber adopted here is a metal-dielectric-metal (MDM) metasurface design based on the standing wave model \cite{nath2015far, to2020detailed, zhou2021ultra, liu2010infrared}. The plasmonic absorber unit cell structure and dimensions are shown in \textbf{Figure 2} (a). The absorber is composed of periodic disk-shaped titanium nanoantennas and a bottom titanium layer, separated by a germanium dielectric layer. Titanium is chosen as both the nanoantenna and the bottom layer material due to its lossy and reflective properties in the LWIR, resulting in broadband absorption \cite{ding2016broadband}. Germanium is chosen as the dielectric layer due to its LWIR transparency and high refractive index (\(n=4\)). The nanoantenna diameter, unit cell dimension, and layer thicknesses are optimized using finite element simulations to enhance broadband LWIR absorption. \textbf{Figure 2} (b) shows an SEM image of the fabricated spintronic Poisson bolometer device. The top contact strip above the MTJ is covered with the plasmonic absorber. The inset shows the plasmonic nanoantenna array and the MTJ nanopillar before the nanoantenna array is deposited. \textbf{Figure 2} (c) shows the simulated and measured absorptance of the plasmonic absorber. This design enables broadband LWIR absorption higher than 60\% across the entire LWIR spectrum and peaks around 11 \textmu m wavelength. The simulated blackbody spectral radiance at 300 K suggests that the blackbody radiation peaks at the 7-14 \textmu m LWIR spectrum (red shaded region). The absorption of the plasmonic nanoantenna array peaks at the same region to maximize the temperature increase in the sensing layer of the MTJ. Heat transfer simulations performed in COMSOL (see Supporting Information) indicate that the broadband LWIR plasmonic absorber enables around 4 times the temperature increase in the sensing layer compared to previously reported designs \cite{bauer2025exploiting}.

\begin{figure}[ht!]
\centering
  \includegraphics[width=0.95\linewidth]{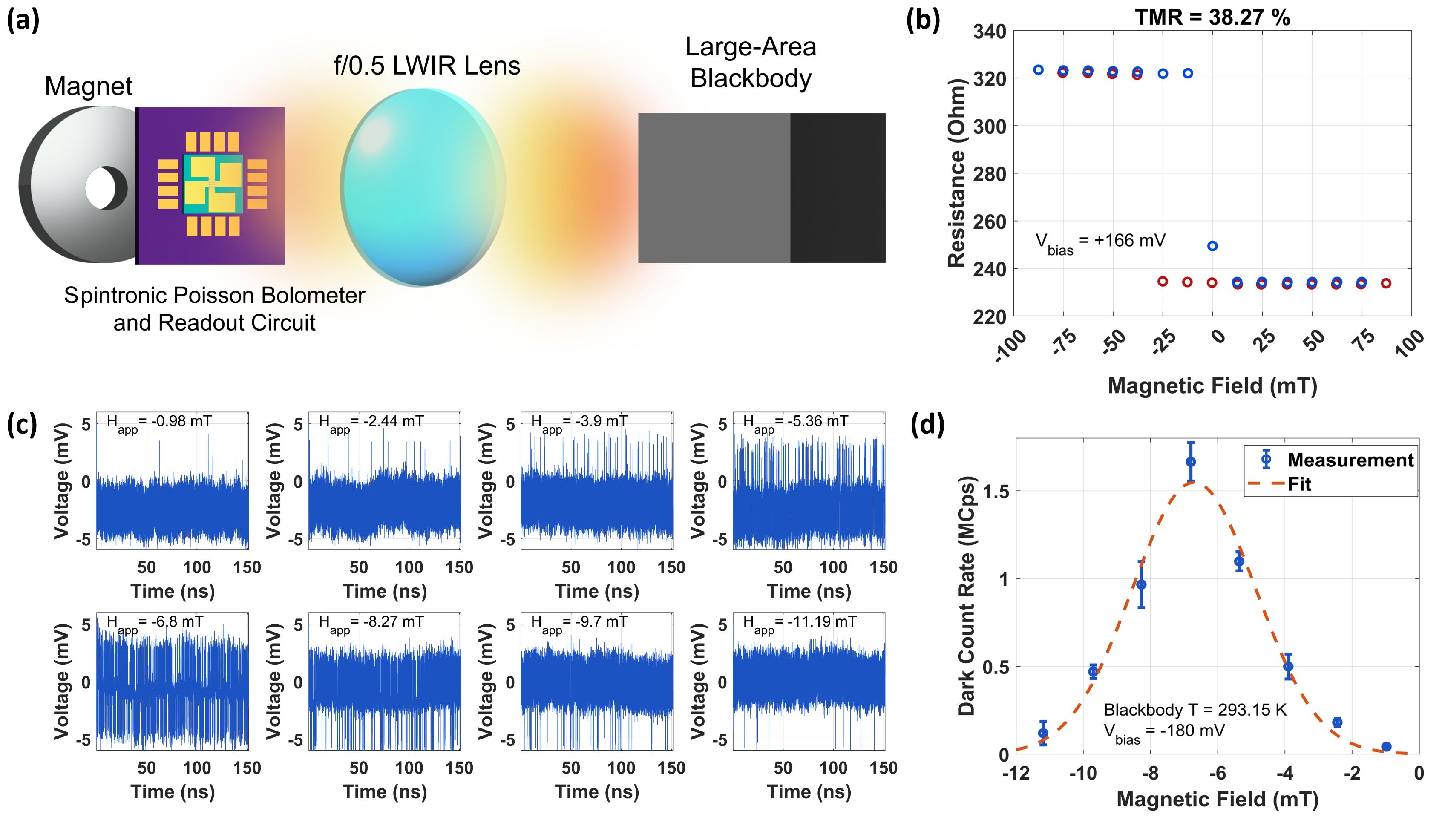}
  \caption{NEDT measurement setup and baseline count rate measurement. (a) NEDT measurement setup schematic. The spintronic Poisson bolometer is wire-bonded to a printed circuit board (PCB), which is connected to readout electronics (see Supporting Information). An electromagnet is placed 8 mm behind the PCB to apply an external bias field $H_{app}$ perpendicular to the MTJ. A ZnSe LWIR lens (f/0.5) is placed at 1.5 cm in front of the PCB board to focus blackbody radiation onto the device. The 18 cm by 18 cm large-area blackbody is placed at 5 cm from the lens and set to different temperatures during NEDT measurement. (b) Tunnel magnetoresistance (TMR) measurement of this spintronic Poisson bolometer device. (c) Voltage waveforms of the baseline transitions when the blackbody temperature is set to 293.15 K (temperature of the environment). A continuous tuning of the count rate by the applied field is observed. (d) Field dependence of the baseline count rate with 20 ms integration time. The dashed line suggests a good fit to the Néel-Arrhenius law.
}
  \label{fig3}
\end{figure}

We first measure the tunnel magnetoresistance (TMR) and field-dependent baseline count rate of the spintronic Poisson bolometer. \textbf{Figure 3} (a) shows the NEDT measurement setup schematic. In \textbf{Figure 3} (b), we find that the TMR is as high as 38\%, which enables high voltage contrast between the two different magnetization states, leading to reliable readout of the count rates. We first set the blackbody temperature to the temperature of the environment (293.15K). We characterize the baseline transitions, i.e., dark counts. \textbf{Figure 3} (c) shows several voltage waveforms, with clear transitions between two voltage levels. We find that the count rate can be tuned by modifying the applied magnetic field. We take a histogram of the waveform and set voltage thresholds accordingly to calculate the number of counts or count rates. \textbf{Figure 3} (d) shows the field dependence of the baseline count rate with 20 ms integration time. The dashed line suggests a good fit to the Néel-Arrhenius law with an offset in the peak count rate from $H = 0$ due to stray fields induced in the sensing layer by the nearby readout and SAF layer. Since the spintronic Poisson bolometer operates via discrete transition events, its dominant noise source is dark counts. Therefore, the ability to control dark counts by modifying the applied magnetic field enables a direct mechanism for tuning and engineering device performance.


We benchmark the setup by measuring the NEDT of a commercial uncooled LWIR detector (FLIR A325sc) based on VOx microbolometers. The NEDT is defined as:

\begin{equation}
NEDT = \frac{C_n}{dC/dT}
\end{equation}

\noindent where $C_n$ is the standard deviation of the count rate at a fixed black body temperature, and $dC/dT$ is the change in count rate per kelvin change in blackbody temperature. \textbf{Figure 4} (a) shows the NEDT result of the VOx microbolometer-based LWIR camera. We measure an NEDT of 79 mK, close to the datasheet's 50 mK sensitivity specification. We compare the FLIR NEDT to five NEDT measurements of a spintronic Poisson bolometer on five different days at different biasing points. The results are shown in \textbf{Figure 4} (b)-(f). \textbf{Figure 4} (b) shows the NEDT result measured for 0.3 K steps in the blackbody temperature. The device shows a linear response between 24.7 and 25.3 \degree C. Afterward, the count rate stops increasing due to device saturation. In \textbf{Figure 4} (c), we repeat the same measurement as \textbf{Figure 4} (b) but with a smaller step size in the linear response region. We find that in this temperature range, the NEDT has improved to 35 mK. In \textbf{Figure 4} (d), we demonstrate the effect of the magnetic field and voltage bias on the NEDT. We find that the count rate decreases and the NEDT increases. The reduced count rate can be explained by current-induced spin-transfer torque and magnetic field dependence, as has been explored previously \cite{bauer2025exploiting}. With proper biasing of the voltage and magnetic field, the spintronic Poisson bolometer can achieve NEDT below 100 mK in multiple measurements on different days. \textbf{Figure 4} (e)-(f) show NEDT measured with negative bias voltage and larger NEDT. This demonstrates that the device sensitivity can be tuned by adjusting the bias voltage and the applied magnetic field.

\begin{figure}[t!]
\centering
  \includegraphics[width=0.95\linewidth]{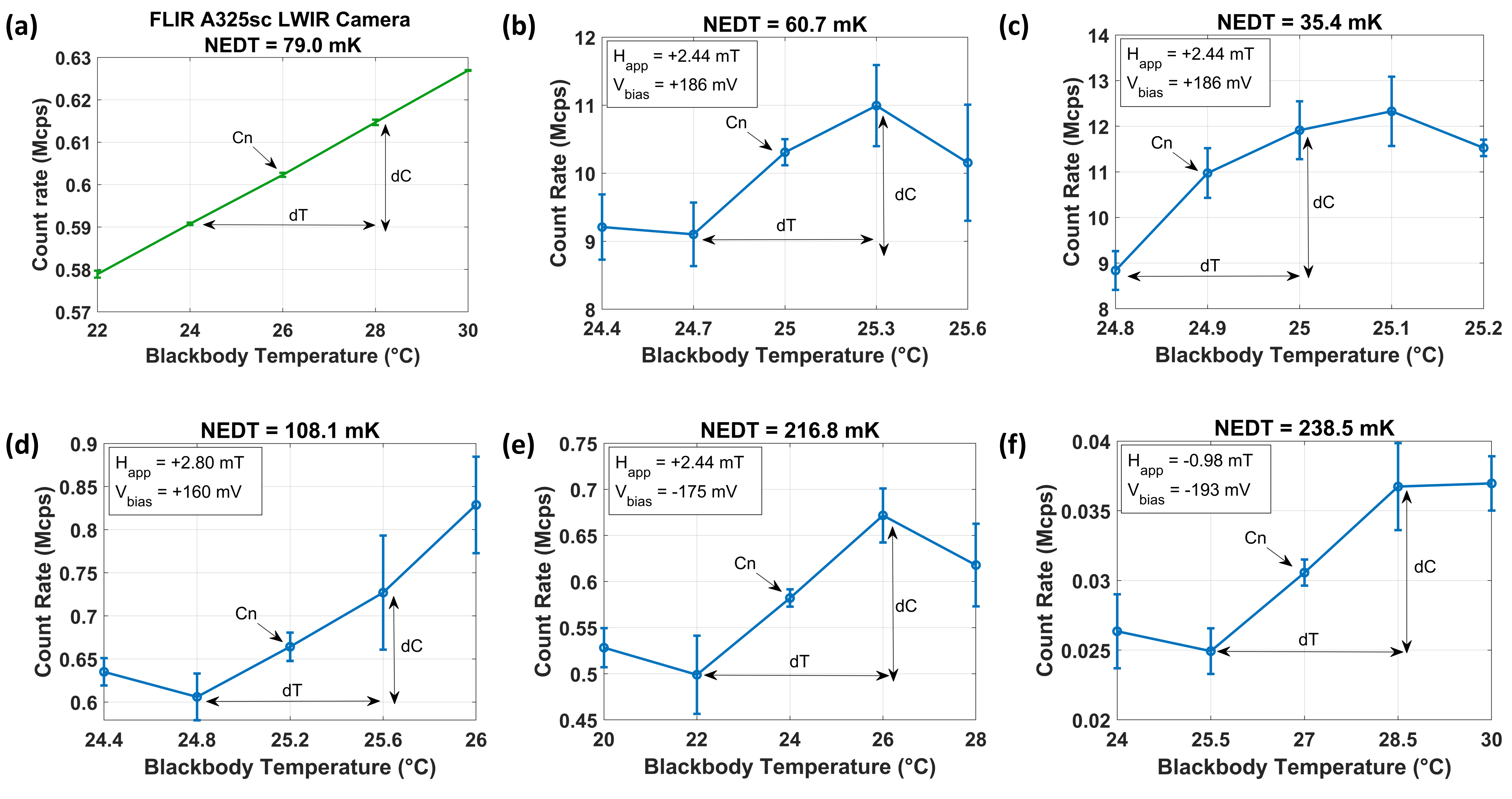}
  \caption{NEDT measurement results. (a) Measured NEDT of a VOx microbolometer-based  LWIR camera (FLIR A325sc). We measure an NEDT of 79 mK, close to the datasheet's 50 mK sensitivity specification. (b) NEDT result measured for 0.3 K steps in the blackbody temperature. For this applied field and bias voltage, the device shows a linear response between 24.7 and 25.3 \degree C. Afterward, the count rate stops increasing due to saturation. (c) Repeated measurement as \textbf{Figure 4} (b) but with a smaller step size in the linear response region. The result shows an improved NEDT of 35 mK. (d) Repeated measurement with decreased bias voltage and slightly increased applied field. The result shows a decreased count rate and higher NEDT due to both the current-induced spin-transfer torque effect and magnetic field dependence. With proper biasing of the voltage and magnetic field, the spintronic Poisson bolometer can achieve NEDT below 100 mK in multiple measurements on different days. (e)-(f) NEDT measured with negative bias voltage. The integration time for (b) - (d) is 20 ms (50 Hz frame rate). The integration time for (e) and (f) is 40 ms (25 Hz frame rate), all comparable to the frame rate of state-of-the-art commercial LWIR cameras.
}
  \label{fig4}
\end{figure}

 The variation of NEDTs in \textbf{Figure 4} (b)-(f) is likely caused by two factors, bias-dependence and measurement limitations. As we previously mentioned, changes to the current and magnetic field affect the count rate, thereby affecting the device's sensitivity \cite{bauer2025exploiting}. Additionally, the limited linear range in some of the measurements can lead to reduced sensitivity measurements, as some temperatures fall outside the linear response range. We also find that the count rate plateaus at higher temperatures, which may be caused by count rate saturation or due to temperature-dependent resistance states. Continued absorption of thermal radiation may cause the energy in the sensing layer to completely surpass the energy barrier, causing the count rate to saturate. Increases in temperature beyond this saturation point would no longer produce a significant rise in the count rate. Alternatively, the number of resistance states, or the relative distance between resistance states, may change with device temperature. Our counting algorithm assumes a strictly two-state response with a fixed relative distance between the two resistance states. Therefore, it may overlook changes in the number or distance of resistance states. Consequently, the spintronic Poisson bolometer responds linearly in a limited temperature range. However, we find that the sensitivity within this linear regime rivals the sensitivity of a commercial uncooled LWIR camera. In \textbf{Table 1}, we compare the thermal sensitivities of an uncooled commercial LWIR camera (FLIR A325sc, VOx microbolometer) and the spintronic Poisson bolometer. To compare the NEDT results with different f-numbers, we normalize the NEDT to f/1.0 using \cite{rogalski2019infrared}:

\begin{equation}
NEDT(f/1.0) = \frac{NEDT(f/\#)}{(f/\#)^2}
\end{equation}

\begin{table}[h!]
\centering
\caption{Comparison of the measured NEDT}
\begin{tabular}[htbp]{@{}c c c c@{}}
\hline
  \textbf{Device} & \textbf{NEDT} & \textbf{Normalized NEDT} & \textbf{Frame Rate} \\
   \hline

   FLIR A325sc (VOx Bolometer) & 79 mK (f/0.6)  & 219.4 mK (f/1.0)  & 30 Hz  \\
   Spintronic Poisson Bolometer & 35.4 mK (f/0.5)  & 141.6 mK (f/1.0) & 50 Hz  \\
   
\hline  
 \end{tabular}
\end{table}

Many high-sensitivity uncooled LWIR detectors are characterized in vacuum chambers \cite{das2023thermodynamically, adiyan2019shape, chen2021ultrafast}, which effectively reduces thermal convection and thereby maximizes the temperature rise induced by incident radiation. We expect that the sensitivity presented in this work could be further enhanced under similar vacuum conditions. The sensitivity achieved in this work establishes spintronic Poisson bolometers as a promising platform for practical applications, such as neuromorphic computing, event-based sensing, and thermal spectral imaging \cite{mousa2025neural, mousa2026uncooled, mousa2026ultra}. Combined with emerging photonic and nanoscale detector technologies, these approaches could enable compact, energy-efficient infrared imaging systems suitable for remote sensing, autonomous navigation, industrial inspection, and rapid thermal event detection.

\section{Conclusion}

In this work, a broadband LWIR plasmonic absorber was integrated onto a spintronic Poisson bolometer to enhance LWIR absorption of the incident radiation and the sensing layer. Through this innovation of integrating spintronic materials and plasmonic materials, the device achieved an NEDT of 35 mK at a 50 Hz frame rate, demonstrating room-temperature performance comparable to the most sensitive uncooled LWIR detectors reported to date. These results represent a significant advancement in the development of uncooled infrared detection technologies, approaching the sensitivity traditionally associated with cryogenic systems. The demonstrated performance highlights the potential of this approach for a wide range of high-sensitivity LWIR applications, including remote thermal monitoring, high-speed imaging, cryogenic system diagnostics, and space-based infrared sensing.

\section{Experimental Section}

\subsection{NEDT Measurement Settings}

For all measurements in this paper, the standard deviation is taken over 5 frames. This NEDT of the FLIR camera is measured with an f/0.6 camera lens and a 30 Hz frame rate. The distance between the camera lens and the blackbody is 5 cm. No external LWIR lens is used for the camera.

\bigskip
\textbf{Supporting Information} \par 
Supporting Information is available from the Wiley Online Library or from the author.

\medskip
\textbf{Acknowledgements} \par 
This work is partially supported by the Elmore Chaired Professorship of Purdue University.

\medskip
\textbf{Conﬂict of Interest} \par The authors declare no conﬂict of interest.


\bibliographystyle{MSP}

\bibliography{ref_all}

\end{document}